\documentclass[11pt]{article}
\usepackage{amsmath,amssymb,amsfonts,bbm}
\usepackage{amsmath,amssymb,amsfonts,bbm}
\usepackage{multirow}
\usepackage[table,xcdraw]{xcolor}
\usepackage{graphicx}
\usepackage{caption}
\usepackage[font={scriptsize}]{subcaption}
\usepackage[export]{adjustbox}
\usepackage{setspace}
\usepackage{float}
\usepackage{adjustbox,lipsum,xcolor}

\newcommand{\be}{\begin{equation}}
\newcommand{\ee}{\end{equation}}
\newcommand{\bea}{\begin{eqnarray}}
\newcommand{\eea}{\end{eqnarray}}
\newcommand{\ba}{\begin{array}}
\newcommand{\ea}{\end{array}}

\newcommand{\htwo}{h_{2,1}}
\newcommand{\M}{\mathcal{M}}
\newcommand{\N}{\mathcal{N}}
\newcommand{\D}{\mathcal{D}}
\newcommand{\K}{\mathcal{K}}

\long\def\symbolfootnote[#1]#2{\begingroup%
\def\thefootnote{\fnsymbol{footnote}}\footnote[#1]{#2}\endgroup}

\newcommand{\RNum}[1]{\uppercase\expandafter{\romannumeral #1\relax}}

\begin{document}

\thispagestyle{empty}\vspace{40pt}

\hfill{}

\vspace{128pt}

\begin{center}
    \textbf{\large The Implications of $\N=2$ Supergravity Cosmology \\On the Topology of the Calabi-Yau Manifold}\\
  \vspace{40pt}

Safinaz Salem$^{a}$\symbolfootnote[1]{\tt safinaz.salem@azhar.edu.eg},
    Moataz H. Emam$^{b,\, c}$\symbolfootnote[2]{\tt moataz.emam@cortland.edu}, and H. H. Salah$^{a}$\symbolfootnote[3]{\tt halahashem.519@azhar.edu.eg} 
\end{center}

    \vspace{3pt}
    \begin{changemargin}{0.5in}{0in}
    \begin{flushleft}
        \begin{spacing}{1.0}
    $^a$ \textit{\small  Department of Physics, Faculty of Science, Al Azhar University, Cairo, Egypt}\\ 
    $^b$ \textit{\small  Department of Physics, SUNY College at Cortland, Cortland, New York 13045, USA}\\
    $^c$ \textit{\small  University of Science and technology, Zewail City of Science and Technology, Giza 12578, Egypt}\\
        \end{spacing}
    \end{flushleft}
    \end{changemargin}

\vspace{6pt}

\begin{abstract}
   When $\N=1$ $D=11$ supergravity is compactified on CY threefold to $\N=2$ $D=5$ supergravity 
the action of the last is given in terms of the geometery of the CY manifold space, namely, in terms of 
the hypermultiplets. There are $z^i(i=1,...,h^{2,1})$ complex structure moduli in the moduli space of the CY manifold which's a special K\"{a}hler  manifold with a metric $G_{i\bar{j}}$. 
We solve the field equations of the complex structure moduli with the solution of the Einstein field equations to the moduli velocity norm $G_{i\bar{j}} z^i z^{\bar{j}}$ in case of a 3- brane filled with radiation, dust and energy
embedded in the bulk of $D=5$ supergravity. We get the time dependence
of the moduli and the metric. Then we can further deduce the geometry of the moduli space 
by getting the K\"{a}hler  potential which directly relates to the volume of the CY manifold.    
\end{abstract}

\newpage



\vspace{15pt}

\pagebreak

\section{Introduction}

The compactification of string theory over Calabi-Yau manifolds yields two sets of parameters \cite{Douglas:2015, Greene:1996cy}. 
The parameters corresponding to the structure of Calabi-Yau manifold $\M$ as a complex manifold and the deformation of
the K\"{a}hler metric of the complex structure space. And parameters corresponding to the deformation of $\M$ as a complex K\"{a}hler manifold $\M_K$.

Calabi-Yau 3-folds admit $H^3$ homolgy group that can be Hodge decomposed as 
\be%
H^3 = H^{3,0} \oplus H^{2,1} \oplus H^{1,2} \oplus H^{0,3}. 
\ee%
So CY 3-folds have a single $(3,0)$ cohomology form, where the hodge number $h_{(3,0)}= \text{dim} (H^{3,0})=1$. 
We will call the holomorphic volume form as $\Omega$, (2,1) forms related to $\M_C$, with a Hodge number $h_{2,1}$ determines
the dimensions of $\M_C$, and (1,2) forms related to $\M_K$, with a Hodge number $h_{1,2}$ determines
the dimensions of $\M_K$.

The deformation of $\M$ can be done by either the deformation of $\M_C$ or the deformation of the 
K\"{a}hler form K of $\M_k$ or both. The K\"{a}hler form is defined by 
\be%
K = i g_{m\bar{n}} d \omega^m \wedge d \omega^{\bar{n}}.
\ee%
The holomorphic coordinates $i,\bar{j}=1,...,m$, where 2m is the dimension of the manifold. The CY metric is defined by
\be%
g_{m\bar{n}} = \partial_m \partial_{\bar{n}} \kappa,
\ee%
where $\kappa$ is the K\"{a}hler potential. This paper is devoted to explore the space of the complex structure moduli $\M_C$ that
is described by the $(2,1)$ forms
\be%
\chi_{i|mn\bar{p}} = - \Omega_{mn}^{\bar{r}} \left(\frac{\partial g_{\bar{p}\bar{r}}}{\partial z^i}\right),
\ee%
where $(z^i: i =1,...,h_{2,1})$ are the parameters or the moduli of the complex structure space. Each $\chi_i$ defines a $(2,1)$
cohomology class. 
It's important here to declare that $z^i$ can be treated as complex coordinates that define a K\"{a}hler metric
$G_{i\bar{j}}$ on $\M_C$ as follows:
\be%
V_{CY}~ G_{i\bar{j}} (\delta z^i) (\delta z^{\bar{j}}) = \frac{1}{4}~ \int_\M~ g^{m\bar{n}}~ g^{r\bar{p}}~ (\delta g_{mr})
~ (\delta g_{\bar {n} \bar {p}}),
\ee%
where $V_{CY}$ is the CY volume. $G_{i\bar{j}}$ is related to the K\"{a}hler potential by 
\be%
G_{i\bar{j}} = \partial_i \partial_{\bar{j}} \kappa
\label{pot1}
\ee%
that leads to a relation between $\kappa$ and the volume form 
\be%
\int_\M \Omega \wedge \bar{\Omega} = -i e^{-\kappa},
\ee%
gives that the K\"{a}hler potential is related to the volume of CY manifold simply by \cite{Candelas:356, Emam:2010}
\be 
\text{Vol} (\M)= e^{-\kappa}
\label{vol}
\ee
We will consider here the Hodge number $h_{2,1} =1$, i.e., we have only one moduli $z$, a single K\"{a}hler metric component 
$G$ and the dimension of $\M_C$ is unity. In this work we aim to find the time dependence of the scalar quantity $G_{i\bar{j}} \dot z^i \dot z^{\bar{j}}$ and then to find:
\begin{itemize}
\item The variation of the moduli $z$ and the K\"{a}hler metric  $G$ with time.
\item The the time dependence of the K\"{a}hler potential $\kappa$ and the Vol($\M$).
\end{itemize}
Our study is based on $D=5$ $\N=2$ supergravity where the universe is modeled as 
a 3-brane embedded in a 5 dimensional bulk. Previously \cite{Emam:2020oyb} we have found that the moduli's velocity norm 
$G_{i\bar{j}} \dot z^i \dot z^{\bar{j}}$ correlates to the scale factors of the brane universe or the bulk  and 
significantly corresponds to our own universe cosmological time evolution. We have studied a 3-brane filled by radiation as our very earl universe and a brane filled by dust, where the Friedmann-like equations have been numerically solved for a wide rang of the scale factors initial conditions. In all different cases $G_{i\bar{j}} \dot z^i \dot z^{\bar{j}}$  manifested it self as an agent
starts with very large values causes an early epoch of rapid expansion (inflation) then it decays fastly to asymptotic values. 
Here, we will extend that study and add to the Einstein's equations a cosmological constant term. We will solve the field equations in case of a 3-brane filled by radiation, dust and energy (cosmological constant $\Lambda$ ), while we consider the bulk's   
cosmological constant $\tilde{\Lambda}$ vanishes. 
Then we will use these results to explore the non- trivial topology of the Calabi-Yau manifold. 
We use the system of units $M_p=1 (M_p = 2.4 \times 10^{18} ~ \text{GeV}= (8 \pi G)^{-1/2}) $.  

It's worthy to mention that there are many studies about the geometry of the moduli spaces for a Calabi-Yau manifold like \cite{Candelas:359, Candelas:355, Belavin:2018, Belavin:2017}, where in \cite{Candelas:359} for instance CY 3-fold was considered as a quintic threefold in the $\mathbb{P}^4$ projection space with $h_{2,1} =101$. However we don't need here to make this assumption.~

So this paper is organized as follows: in section 2 we introduce the $D=5$ $\N=2$ supergravity as 
the dimensional reduction of eleven dimensional supergravity theory over a Calabi-Yau 3-fold $\M$.
Then we will solve the field equations.  
In section 3 we will simplify the moduli field equations and the K\"{a}hler metric equation and 
show how they can be solved using the solution of the moduli's velocity norm 
$G_{i\bar{j}} \dot z^i \dot z^{\bar{j}}$. 
In the same section we will introduce the time dependence of the K\"{a}hler metric, K\"{a}hler potential and Vol ($\M$).

\section{$\D=5$ $\N=2$ supergravity and its solution}

The five dimensional $\N=2$ supergravity theory contains two sets of matter fields; the vector multiplets, which we set to zero, and our main interest: the \emph{hypermultiplets}. These are composed of the \emph{universal hypermultiplet} $\left(\phi, \sigma, \zeta^0, \tilde \zeta_0\right)$; where $\phi$ is the universal axion, and the dilaton $\sigma$ is proportional to the volume of the underlying Calabi-Yau manifold $\M$. The remaining hypermultiplet scalars are $\left(z^i, z^{\bar i}, \zeta^i, \tilde \zeta_i: i=1,\ldots, \htwo\right)$, where the $z$'s are the complex structure moduli of $\M$, and $\htwo$ is the Hodge number determining the dimensions of the manifold $\M_C$ of the Calabi-Yau's complex structure moduli\footnote{A `bar' over an index denotes complex conjugation}. The fields $\left(\zeta^I, \tilde\zeta_I: I=0,\ldots,\htwo\right)$ are the axions, which define a symplectic vector space (see \cite{Emam:2010} for a review and more references). The axions are defined as components of the symplectic vector
\be\label{DefOfSympVect}
   \left| \Xi  \right\rangle  = \left( {\begin{array}{*{20}c}
   {\,\,\,\,\,\zeta ^I }  \\
   -{\tilde \zeta _I }  \\
    \end{array}} \right),
\ee
such that the symplectic scalar product is defined by, for example,
\be
    \left\langle {{\Xi }}
 \mathrel{\left | {\vphantom {{\Xi } \Xi }}
 \right. \kern-\nulldelimiterspace}
 {\Xi } \right\rangle   = \zeta^I \tilde \zeta_I  - \tilde \zeta_I
 \zeta^I.\label{DefOfSympScalarProduct}
\ee

A transformation in symplectic space can be defined by
\be
 \left\langle {d\Xi } \right|\mathop {\bf\Lambda} \limits_ \wedge  \left| {\star d\Xi } \right\rangle
  = 2\left\langle {{d\Xi }}
 \mathrel{\left | {\vphantom {{d\Xi } V}}
 \right. \kern-\nulldelimiterspace}
 {V} \right\rangle \mathop {}\limits_ \wedge  \left\langle {{\bar V}}
 \mathrel{\left | {\vphantom {{\bar V} {\star d\Xi }}}
 \right. \kern-\nulldelimiterspace}
 {{\star d\Xi }} \right\rangle  + 2G^{i\bar j} \left\langle {{d\Xi }}
 \mathrel{\left | {\vphantom {{d\Xi } {U_{\bar j} }}}
 \right. \kern-\nulldelimiterspace}
 {{U_{\bar j} }} \right\rangle \mathop {}\limits_ \wedge  \left\langle {{U_i }}
 \mathrel{\left | {\vphantom {{U_i } {\star d\Xi }}}
 \right. \kern-\nulldelimiterspace}
 {{\star d\Xi }} \right\rangle  - i\left\langle {d\Xi } \right.\mathop |\limits_ \wedge  \left. {\star d\Xi } \right\rangle,\label{DefOfRotInSympSpace}
\ee
where $d$ is the spacetime exterior derivative, $\star$ is the five dimensional Hodge duality operator, and $G_{i\bar j}$ is a special K\"{a}hler metric on $\M_C$. The symplectic basis vectors $\left| V \right\rangle $, $\left| {U_i } \right\rangle $ and their complex conjugates are defined by
\be
    \left| V \right\rangle  = e^{\frac{\K}{2}} \left( {\begin{array}{*{20}c}
   {Z^I }  \\
   {F_I }  \\
    \end{array}} \right),\,\,\,\,\,\,\,\,\,\,\,\,\,\,\,\left| {\bar V} \right\rangle  = e^{\frac{\K}{2}} \left( {\begin{array}{*{20}c}
   {\bar Z^I }  \\
   {\bar F_I }  \\
    \end{array}} \right)\label{DefOfVAndVBar}
\ee

\noindent where $\K$ is the K\"{a}hler potential on $\M_C$, $\left( {Z,F} \right)$ are the periods of the Calabi-Yau's holomorphic volume form, and

\bea
    \left| {U_i } \right\rangle  &=& \left| \nabla _i V
    \right\rangle=\left|\left[ {\partial _i  + \frac{1}{2}\left( {\partial _i \K} \right)} \right] V \right\rangle \nonumber\\
    \left| {U_{\bar i} } \right\rangle  &=& \left|\nabla _{\bar i}  {\bar V} \right\rangle=\left|\left[ {\partial _{\bar i}  + \frac{1}{2}\left( {\partial _{\bar i} \K} \right)} \right] {\bar V}
    \right\rangle\label{DefOfUAndUBar}
\eea
where the derivatives are with respect to the moduli $\left(z^i, z^{\bar i}\right)$. In this language, the bosonic part of the action is given by:
\bea
    S_5  &=& \int\limits_5 {\left[ {R\star \mathbf{1} - \frac{1}{2}d\sigma \wedge\star d\sigma  - G_{i\bar j} dz^i \wedge\star dz^{\bar j} } \right.}  + e^\sigma   \left\langle {d\Xi } \right|\mathop {\bf\Lambda} \limits_ \wedge  \left| {\star d\Xi } \right\rangle\nonumber\\
    & &\left. {\quad\quad\quad\quad\quad\quad\quad\quad\quad\quad\quad\quad\quad - \frac{1}{2} e^{2\sigma } \left[ {d\phi + \left\langle {\Xi } \mathrel{\left | {\vphantom {\Xi  {d\Xi }}} \right. \kern-\nulldelimiterspace} {{d\Xi }}    \right\rangle} \right] \wedge \star\left[ {d\phi + \left\langle {\Xi } \mathrel{\left | {\vphantom {\Xi  {d\Xi }}} \right. \kern-\nulldelimiterspace} {{d\Xi }}    \right\rangle} \right] } \right].\label{action}
\eea

The usual $\delta S = 0$ gives the following field equations for the hypermultiplets scalar fields:
\bea
    \left( {\Delta \sigma } \right)\star \mathbf{1} + e^\sigma   \left\langle {d\Xi } \right|\mathop {\bf\Lambda} \limits_ \wedge  \left| {\star d\Xi } \right\rangle -   e^{2\sigma }\left[ {d\phi + \left\langle {\Xi } \mathrel{\left | {\vphantom {\Xi  {d\Xi }}} \right. \kern-\nulldelimiterspace} {{d\Xi }}    \right\rangle} \right]\wedge\star\left[ {d\phi + \left\langle {\Xi } \mathrel{\left | {\vphantom {\Xi  {d\Xi }}} \right. \kern-\nulldelimiterspace} {{d\Xi }}    \right\rangle} \right] &=& 0\label{DilatonEOM}\\
    \left( {\Delta z^i } \right)\star \mathbf{1} + \Gamma _{jk}^i dz^j  \wedge \star dz^k  + \frac{1}{2}e^\sigma  G^{i\bar j}  {\partial _{\bar j} \left\langle {d\Xi } \right|\mathop {\bf\Lambda} \limits_ \wedge  \left| {\star d\Xi } \right\rangle} &=& 0 \nonumber\\
    \left( {\Delta z^{\bar i} } \right)\star \mathbf{1} + \Gamma _{\bar j\bar k}^{\bar i} dz^{\bar j}  \wedge \star dz^{\bar k}  + \frac{1}{2}e^\sigma  G^{\bar ij}  {\partial _j \left\langle {d\Xi } \right|\mathop {\bf\Lambda} \limits_ \wedge  \left| {\star d\Xi } \right\rangle}  &=& 0\label{ZZBarEOM} \\
    d^{\dag} \left\{ {e^\sigma  \left| {{\bf\Lambda} d\Xi } \right\rangle  - e^{2\sigma } \left[ {d\phi + \left\langle {\Xi }
    \mathrel{\left | {\vphantom {\Xi  {d\Xi }}}\right. \kern-\nulldelimiterspace} {{d\Xi }} \right\rangle } \right]\left| \Xi  \right\rangle } \right\} &=& 0\label{AxionsEOM}\\
    d^{\dag} \left[ {e^{2\sigma } d\phi + e^{2\sigma } \left\langle {\Xi } \mathrel{\left | {\vphantom {\Xi  {d\Xi }}} \right. \kern-\nulldelimiterspace} {{d\Xi }}    \right\rangle} \right] &=&    0\label{aEOM}
\eea
where $d^\dagger$ is the $D=5$ adjoint exterior derivative, $\Delta$ is the Laplace-de Rahm operator and $\Gamma _{jk}^i$ is a connection on $\M_C$. The full action is symmetric under the following SUSY transformations:
\bea
 \delta _\epsilon  \psi ^1  &=& D \epsilon _1  + \frac{1}{4}\left\{ {i {e^{\sigma } \left[ {d\phi + \left\langle {\Xi }
 \mathrel{\left | {\vphantom {\Xi  {d\Xi }}}
 \right. \kern-\nulldelimiterspace} {{d\Xi }} \right\rangle } \right]}- Y} \right\}\epsilon _1  - e^{\frac{\sigma }{2}} \left\langle {{\bar V}}
 \mathrel{\left | {\vphantom {{\bar V} {d\Xi }}} \right. \kern-\nulldelimiterspace} {{d\Xi }} \right\rangle\epsilon _2  \nonumber\\
 \delta _\epsilon  \psi ^2  &=& D \epsilon _2  - \frac{1}{4}\left\{ {i {e^{\sigma } \left[ {d\phi + \left\langle {\Xi }
 \mathrel{\left | {\vphantom {\Xi  {d\Xi }}} \right. \kern-\nulldelimiterspace}
 {{d\Xi }} \right\rangle } \right]}- Y} \right\}\epsilon _2  + e^{\frac{\sigma }{2}} \left\langle {V}
 \mathrel{\left | {\vphantom {V {d\Xi }}} \right. \kern-\nulldelimiterspace} {{d\Xi }} \right\rangle \epsilon _1,  \label{SUSYGraviton}
\eea
\bea
  \delta _\epsilon  \xi _1^0  &=& e^{\frac{\sigma }{2}} \left\langle {V}
    \mathrel{\left | {\vphantom {V {\partial _\mu  \Xi }}} \right. \kern-\nulldelimiterspace} {{\partial _\mu  \Xi }} \right\rangle  \Gamma ^\mu  \epsilon _1  - \left\{ {\frac{1}{2}\left( {\partial _\mu  \sigma } \right) - \frac{i}{2} e^{\sigma } \left[ {\left(\partial _\mu \phi\right) + \left\langle {\Xi }
    \mathrel{\left | {\vphantom {\Xi  {\partial _\mu \Xi }}} \right. \kern-\nulldelimiterspace}
    {{\partial _\mu \Xi }} \right\rangle } \right]} \right\}\Gamma ^\mu  \epsilon _2  \nonumber\\
     \delta _\epsilon  \xi _2^0  &=& e^{\frac{\sigma }{2}} \left\langle {{\bar V}}
    \mathrel{\left | {\vphantom {{\bar V} {\partial _\mu  \Xi }}} \right. \kern-\nulldelimiterspace} {{\partial _\mu  \Xi }} \right\rangle \Gamma ^\mu  \epsilon _2  + \left\{ {\frac{1}{2}\left( {\partial _\mu  \sigma } \right) + \frac{i}{2} e^{\sigma } \left[ {\left(\partial _\mu \phi\right) + \left\langle {\Xi }
    \mathrel{\left | {\vphantom {\Xi  {\partial _\mu \Xi }}} \right. \kern-\nulldelimiterspace}
    {{\partial _\mu \Xi }} \right\rangle } \right]} \right\}\Gamma ^\mu  \epsilon
     _1,\label{SUSYHyperon1}
\eea
and
\bea
     \delta _\epsilon  \xi _1^{\hat i}  &=& e^{\frac{\sigma }{2}} e^{\hat ij} \left\langle {{U_j }}
    \mathrel{\left | {\vphantom {{U_j } {\partial _\mu  \Xi }}} \right. \kern-\nulldelimiterspace} {{\partial _\mu  \Xi }} \right\rangle \Gamma ^\mu  \epsilon _1  - e_{\,\,\,\bar j}^{\hat i} \left( {\partial _\mu  z^{\bar j} } \right)\Gamma ^\mu  \epsilon _2  \nonumber\\
     \delta _\epsilon  \xi _2^{\hat i}  &=& e^{\frac{\sigma }{2}} e^{\hat i\bar j} \left\langle {{U_{\bar j} }}
    \mathrel{\left | {\vphantom {{U_{\bar j} } {\partial _\mu  \Xi }}} \right. \kern-\nulldelimiterspace} {{\partial _\mu  \Xi }} \right\rangle \Gamma ^\mu  \epsilon _2  + e_{\,\,\,j}^{\hat i} \left( {\partial _\mu  z^j } \right)\Gamma ^\mu  \epsilon    _1,\label{SUSYHyperon2}
\eea
where $\left(\psi ^1, \psi ^2\right)$ are the two gravitini and $\left(\xi _1^I, \xi _2^I\right)$ are the hyperini. The quantity $Y$ is defined by:
\begin{equation}
    Y   = \frac{{\bar Z^I N_{IJ}  {d  Z^J }  -
    Z^I N_{IJ}  {d  \bar Z^J } }}{{\bar Z^I N_{IJ} Z^J
    }},\label{DefOfY}
\end{equation}
where $N_{IJ}  = \mathfrak{Im} \left({\partial_IF_J } \right)$. The $e$'s are the beins of the special K\"{a}hler metric $G_{i\bar j}$, the $\epsilon$'s are the five-dimensional $\N=2$ SUSY spinors and the $\Gamma$'s are the usual Dirac matrices. The covariant derivative $D$ is defined by the usual $D=dx^\mu\left( \partial _\mu   + \frac{1}{4}\omega _\mu^{\,\,\,\,\hat \mu\hat \nu} \Gamma _{\hat \mu\hat \nu}\right)\label{DefOfCovDerivative}$, where the $\omega$'s are the spin connections and the hatted indices are frame indices in a flat tangent space. Finally, the stress tensor is:
\bea
T_{\mu \nu }  &=& -\frac{1}{2}\left( {\partial _\mu  \sigma } \right)\left( {\partial _\nu  \sigma } \right) + \frac{1}{4}g_{\mu \nu } \left( {\partial _\alpha  \sigma } \right)\left( {\partial ^\alpha  \sigma } \right)
 + e^\sigma  \left\langle {\partial _\mu \Xi } \right|{\bf\Lambda} \left| {\partial _\nu \Xi } \right\rangle - \frac{1}{2}e^{\sigma } g_{\mu \nu }   \left\langle {\partial _\alpha \Xi } \right|{\bf\Lambda} \left| {\partial ^\alpha \Xi } \right\rangle \nonumber\\
  & &  - \frac{1}{2}e^{2\sigma } \left[ {\left( {\partial _\mu  \phi} \right) + \left\langle {\Xi }
 \mathrel{\left | {\vphantom {\Xi  {\partial _\mu  \Xi }}}
 \right. \kern-\nulldelimiterspace}
 {{\partial _\mu  \Xi }} \right\rangle } \right]\left[ {\left( {\partial _\nu  \phi} \right) + \left\langle {\Xi }
 \mathrel{\left | {\vphantom {\Xi  {\partial _\nu  \Xi }}}
 \right. \kern-\nulldelimiterspace}
 {{\partial _\nu  \Xi }} \right\rangle } \right]
  +  \frac{1}{4}e^{2\sigma } g_{\mu \nu } \left[ {\left( {\partial _\alpha  \phi} \right) + \left\langle {\Xi }
 \mathrel{\left | {\vphantom {\Xi  {\partial _\alpha  \Xi }}}
 \right. \kern-\nulldelimiterspace}
 {{\partial _\alpha  \Xi }} \right\rangle } \right]\left[ {\left( {\partial ^\alpha  \phi} \right) + \left\langle {\Xi }
 \mathrel{\left | {\vphantom {\Xi  {\partial ^\alpha  \Xi }}}
 \right. \kern-\nulldelimiterspace}
 {{\partial ^\alpha  \Xi }} \right\rangle } \right]\nonumber\\
 & & - G_{i\bar j} \left( {\partial _\mu  z^i } \right)\left( {\partial _\nu  z^{\bar j} } \right) + \frac{1}{2}g_{\mu \nu } G_{i\bar j} \left( {\partial _\alpha  z^i } \right)\left( {\partial ^\alpha  z^{\bar j} } \right).\label{StressTensor}
\eea

\begin{figure}[H]
 \centering
    \includegraphics[width=0.5\columnwidth]{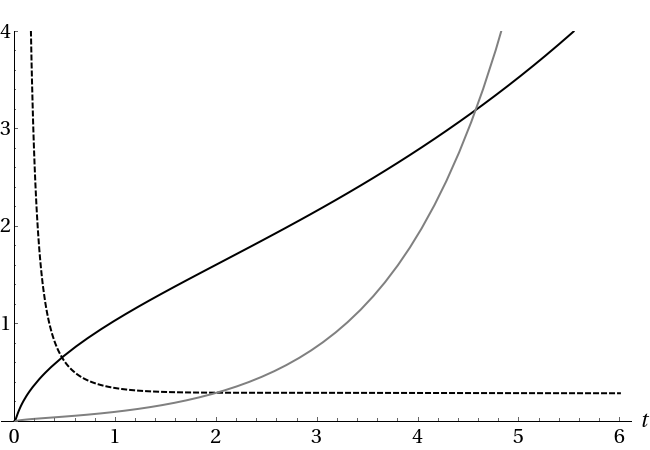}
    \caption{The scale factor $a$ is represented by the solid curve, $b$ by the grey curve, while $\left| {G_{i\bar j} \dot z^i \dot z^{\bar j}} \right|$ is shown dashed. $\Lambda=1, \tilde{\Lambda}=0$, and for initial conditions
$a[0]=a'[0]=b[0]=b'[0] = 0$ .}
    \label{abzz}
  \end{figure}

As our main interest is bosonic configurations that preserve \emph{some} supersymmetry, the stress tensor can be simplified by considering the vanishing of the supersymmetric variations (\ref{SUSYHyperon1},\ref{SUSYHyperon2}); satisfying the BPS condition on the brane. This gives
\be
    T_{\mu \nu }  = G_{i\bar j} \left( {\partial _\mu  z^i } \right)\left( {\partial _\nu  z^{\bar j} } \right) - \frac{1}{2}g_{\mu \nu } G_{i\bar j} \left( {\partial _\alpha  z^i } \right)\left( {\partial ^\alpha  z^{\bar j} } \right),\label{StressTens}
\ee
as was detailed out in \cite{Emam:2015laa}. We would like to construct a 3-brane that may be thought of as a flat Robertson-Walker universe embedded in $D=5$. As such we invoke the metric
\be
    ds^2  =  - dt^2  + a^2 \left( t \right) \left( {dr^2  + r^2 d\Omega ^2 } \right) + b^2 \left( t \right) dy^2,
\label{RWM}
\ee
where ${d\Omega ^2  = d\theta ^2  + \sin ^2 \left( \theta  \right)d\varphi ^2 }$, $a^2 \left( t \right)$ is the usual Robertson-Walker scale factor, and $b \left( t \right)$ is a possible scale factor for the transverse dimension $y$ (the bulk). The brane is located at $y=0$ and we ignore all possible $y$-dependence of the warp factors as well as the hypermultiplet bulk fields; effectively only studying the brane close to its location. In this case Einstein equations $G_{MN} + \Lambda g_{MN} = T_{MN} $ reduce to the Friedmann-like form:
\bea
 3\left[ {\left( {\frac{{\dot a}}{a}} \right)^2  + \left( {\frac{{\dot a}}{a}} \right)\left( {\frac{{\dot b}}{b}} \right)} \right] &=& G_{i\bar j} \dot z^i \dot z^{\bar j}  + \rho+ \Lambda \nonumber\\
-\left[ 2\frac{{\ddot a}}{a} + \left( {\frac{{\dot a}}{a}} \right)^2  + \frac{{\ddot b}}{b} + 2\left( {\frac{{\dot a}}{a}} \right)\left( {\frac{{\dot b}}{b}} \right) \right] &=&  G_{i\bar j} \dot z^i \dot z^{\bar j}  + p - \Lambda \nonumber\\
 -3\left[ {\frac{{\ddot a}}{a} + \left( {\frac{{\dot a}}{a}} \right)^2 } \right] &=&   G_{i\bar j} \dot z^i \dot z^{\bar j}- \tilde{\Lambda}. \label{EF11}
\eea
Solving these equations numerically in case if the total density equals the dust plus radiation densities $\rho = \rho_r+ \rho_m \propto 1/a^4 +1/a^3$, the total pressure equals to the radiation pressure $p=p_r=\rho_r/3$, $\Lambda=1$ (de Sitter space), and 
$\tilde{\Lambda}=0$ yields the  brane's and the bulk's scale factors and $\left| {G_{i\bar j} \dot z^i \dot z^{\bar j}} \right|$
as functions in time Fig. (\ref{abzz}). Using  suitable fitting functions we get the solution of the velocity norm of the complex structure moduli given by
\be
{G_{i\bar j} \dot z^i \dot z^{\bar j}} (t) \simeq -0.5 (t + 0.004)^{-0.9}.
\label{zsol}
\ee
The brane's scale factor varies with time exponentially $a(t) \sim e^{0.2~t}$ which means 
the brane- universe undergoes an inflationary expansion. While the bulk scale factor is given by 
$b(t) \sim 0.06 e^{0.87~t}$.
As seen the solution shows a correlation between the scale factors of the brane universe and the bulk 
and the moduli norm ${G_{i\bar j} \dot z^i \dot z^{\bar j}}$.

\section{Calabi-Yau manifold complex structure space}
The field equations of the moduli $z^i$ and $z^{\bar{i}}$ (\ref{ZZBarEOM})
can be simplified using the BPS condition \cite{Emam:2015laa}:
\be e^\sigma \langle \Xi|\underset{\Lambda}{\bf \Lambda}| \star d\Xi \rangle
 = \frac{1}{2} d\sigma \wedge \star d\sigma + \frac{1}{2} e^{2\sigma} [d\phi + \langle\Xi|~d\Xi \rangle] \wedge \star [d\phi + \langle \Xi|~d\Xi\rangle] + 2 G_{i\bar{j}} dz^i \wedge \star dz^{\bar{j}},\ee
so that they can be written as 
\bea \nonumber
& (\Delta z^i) \star 1 + \Gamma^i_{jk} dz^j \wedge \star dz^k + G^{i\bar{j}} (\partial_{\bar{j}} G_{l\bar{k}}) dz^l \wedge \star dz^{\bar{k}}   =0 \\
& (\Delta z^{\bar{i}}) \star 1 + \Gamma^{\bar{i}}_{\bar{j}\bar{k}} dz^{\bar{j}} \wedge \star dz^{\bar{k}} + G^{\bar{i}j} (\partial_j G_{l\bar{k}}) dz^l \wedge \star dz^{\bar{k}}  =0.
\eea
Dropping the differential forms formulation, we get:
\bea \nonumber
& \nabla^2 z^i + \Gamma^i_{jk} \partial_\mu z^j \partial^\mu z^k + G^{i\bar{j}} (\partial_{\bar{j}} G_{l\bar{k}})\partial_\mu z^l \partial^\mu z^{\bar{k}}  =0 \\
& \nabla^2 z^{\bar{i}} + \Gamma^{\bar{i}}_{\bar{j}\bar{k}} \partial_\mu z^{\bar{j}} \partial^\mu z^{\bar{k}} +
G^{\bar{i}j} (\partial_{j} G_{l\bar{k}})\partial_\mu z^l \partial^\mu z^{\bar{k}}   =0.
\label{zz23}\eea
The connections or the Christoffel symbols are related to the metric by \cite{Emam:2010}:
\be 
\Gamma^i_{jk} = G^{i\bar{p}} \partial_j G_{k\bar{p}}, ~~~~~~~~ \Gamma^{\bar{i}}_{\bar{j}\bar{k}} = G^{p\bar{i}} \partial_{\bar{j}} 
G_{\bar{k}p},
\ee
substitute in (\ref{zz23}), we get:
\bea \nonumber
& \nabla^2 z^i + G^{i\bar{p}} \partial_j G_{k\bar{p}} ~ \partial_\mu z^j \partial^\mu z^k + G^{i\bar{j}} (\partial_{\bar{j}} G_{l\bar{k}})\partial_\mu z^l \partial^\mu z^{\bar{k}}  =0 \\
& \nabla^2 z^{\bar{i}} +  G^{p\bar{i}} \partial_{\bar{j}} 
G_{\bar{k}p}~ \partial_\mu z^{\bar{j}} \partial^\mu z^{\bar{k}} +
G^{\bar{i}j} (\partial_{j} G_{l\bar{k}})\partial_\mu z^l \partial^\mu z^{\bar{k}}   =0 
\label{zz34}\eea
The moduli are independent of the 3- spatial dimensions. And consider the Hodge number $h_{2,1} =1$,  which means we have only one moduli $z$, its complex conjugate $z^*$, a single K\"{a}hler metric component $G$ and the dimension of the moduli space $\M_C$ is unity. So that 
(\ref{zz34}) simplify to:    
\bea \nonumber
&  \ddot{z} +\frac {1}{G} (\partial_z G) ~ \dot {z}^2 + \frac {1}{G} (\partial_{z^*} G) \dot{z} \dot{z}^*  =0 \\
& \ddot{z}^* + \frac {1}{G^*}(\partial_{z^*} G^*) ~ \dot {z^*}^2 + \frac {1}{G^*} (\partial_z G) \dot{z} \dot{z}^* =0 
\eea
From the Robertson-Walker like metric (\ref{RWM}), the moduli field equations become:
\be
\ddot{z} + \frac {1}{G} (\partial_z G) ~ \dot {z}^2 + \frac {1}{G} (\partial_{z^*} G) \dot{z} \dot{z}^*  =0, 
\label{z}
\ee
\be
\ddot{z}^* + \frac {1}{G^*}(\partial_{z^*} G^*) ~ \dot {z^*}^2 + \frac {1}{G^*} (\partial_z G) \dot{z} \dot{z}^* =0. 
\label{zstar}
\ee
Solving the moduli field equation Equ. (\ref{z}) with Equ. (\ref{zsol}), gives
the moduli's variation with time:
\be
z(t) \simeq C + \frac{0.001 +0.25~ t}{(1 + 250~ t)^{0.6}}.
\label{z-t}
\ee
For $\dot{z}[0]=0$. C is the integration constant, for $z[0]=1$, $C \sim 1$. We have made a further approximation here by considering the moduli real. In Fig. (\ref{z}- left) and (\ref{z}- right) the moduli and the moduli velocity are plotted versus time, respectively.

\begin{figure}[H]
\centering
\includegraphics[scale=0.35]{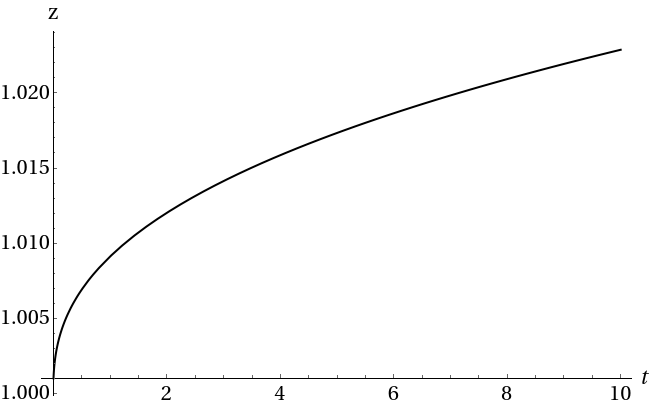}
\includegraphics[scale=0.35]{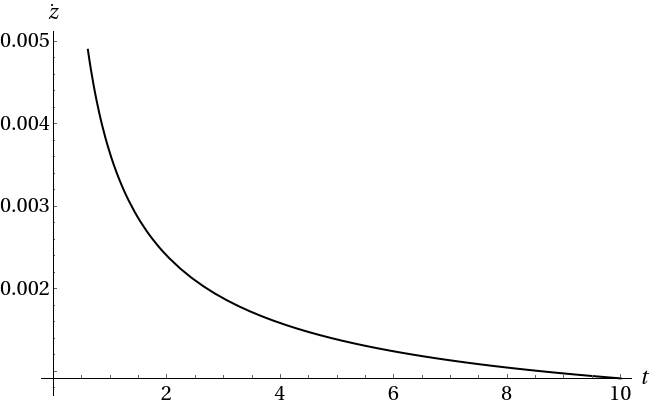}
\caption{(Left panel): The moduli is plotted versus time. 
(Right panel): The moduli velocity is plotted versus time. For $C =1$, and in case 
of radiation, dust and $\Lambda$ filled brane with $\Lambda=1$ and $\tilde{\Lambda}=0$. }
\label{z-zdot}
\end{figure}
\begin{figure}[!h]
\centering
\includegraphics[scale=0.35]{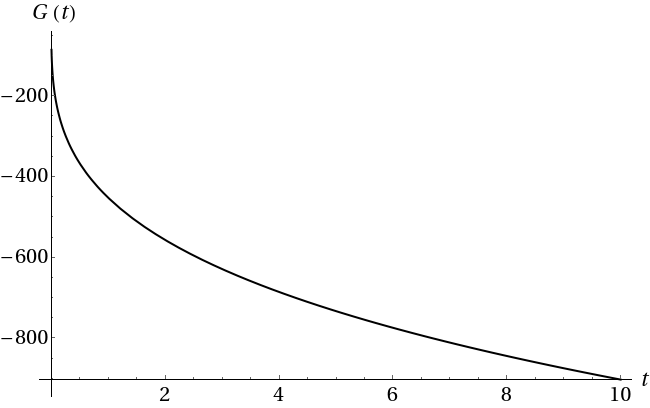}
\includegraphics[scale=0.35]{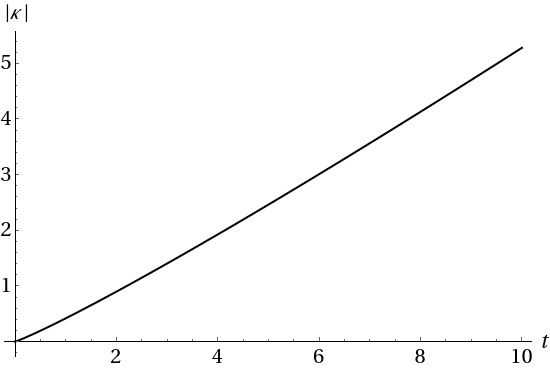}
\caption{(Left panel): One component of the K\"{a}hler  metric multiplied by a factor $10^{-2}$ is
plotted versus time. (Right panel):
The K\"{a}hler potential of $\M_C$ is plotted versus time.}
\label{G}
\end{figure}

The K\"{a}hler metric can be directly obtained by substituting $\dot{z}$ solution in Equ. (\ref{zsol}). In Fig. (\ref{G}- left)
one component of the metric $G_{i\bar{j}}$ multiplied by a factor $10^{-2}$ is plotted versus time. 
Generally speaking the K\"{a}hler potential is given by:
\be 
\kappa= ln ~\left[ 1+ z_i z^{\bar{j}}\right] = ln~\left[ 1+\delta_{i\bar{j}} z^i z^{\bar{j}}\right],
\ee
in which the K\"{a}hler metric Equ. (\ref{pot1}) has been drived.
We will use Equ. (\ref{pot1}) to get $\kappa$ as a function in time.
According to our approximation it can be written as:
\be 
G = \partial_z \partial_{z^*} \kappa.
\label{Gk}
\ee  
Let all fields depend on time, it becomes:
\be
G(t)  = \frac{\partial}{\partial t} \left( \frac{1}{\dot{z}} \frac{\dot{k}}{\dot{z^*}} \right).
\ee
Solving for the K\"{a}hler potential, yields:
\be
\kappa (t) = - \left( 0.45~ t+ 0.008 ~ t^2 \right).
\ee
Fig. (\ref{G}- right) shows the absolute value of the potential plotted versus time. According to
Equ. (\ref{vol}) the volume of the CY manifold $\M$ can be obtained as long as the K\"{a}hler potential is known.
Fig. (\ref{vol}- left) shows the volume of the Calabi-Yau manifold plotted versus time . As seen it increases with time.
\begin{figure}[!t]
\centering
\includegraphics[scale=0.35]{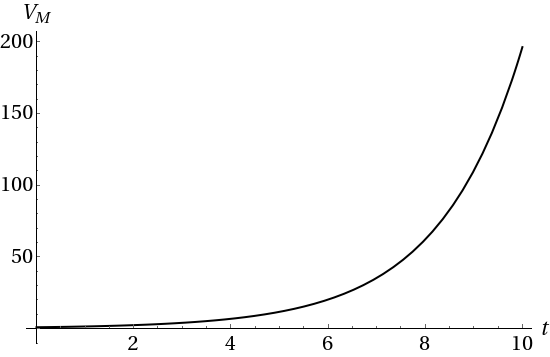}
\includegraphics[scale=0.35]{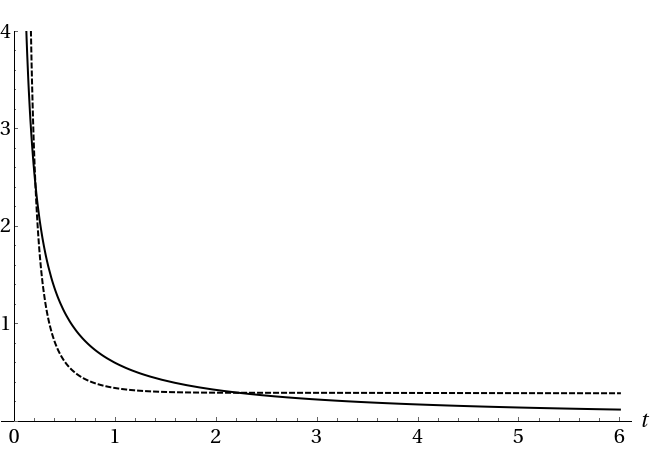}
\caption{(Left panel): The volume of the Calabi-Yau manifold plotted versus time.
The Hodge number $h_{2,1}=1$.
(Right panel):
Numeric and analytic $G_{i\bar{j}} \dot z^i \dot z^{\bar{j}}$ versus time in solid and dashed lines, respectively.
The Hodge number $h_{2,1}=1$.}
\label{vol}
\end{figure}
For the sake of comparison, Fig. (\ref{vol}- right) shows $G_{i\bar{j}} \dot z^i \dot z^{\bar{j}}$ plotted versus time as it's obtained directly from the numeric solution of the field equations (\ref{EF11}) without any approximations a long a side as it's obtained when solving Equ. (\ref{zsol}) with Equ. (\ref{z-t}).

\section{Conclusion}
Exploring the non-trivial topology of the Calabi-Yau manifold still
a highly demanding quest in theoretical physics. The importance of the  
Calabi-Yau manifold arises from its vital role in the compactification of many 
higher dimensional theories. Like when compactifying $\D=11$ supergravity to 
$\N=2$, $\D=5$ supergravity over CY 3-fold. In this work we have studied 
a 3-brane embedded in the bulk of $\D=5$ supergravity, we have solved the  
Friedmann-like equations in case of a brane filled by radiation, dust and energy.
We have shown that in this case the time evolution of the 3-brane coincide with the time evolution 
of our universe, where the moduli's velocity norm $G_{i\bar{j}} \dot z^i \dot z^{\bar{j}}$ strongly 
correlates to the scale factor of the brane-universe. That means the cosmology of our universe can be 
interpreted only by the effects of the bulk of a higher dimensional theory. This explanation needs more 
analysis and initial conditions to be verified which we keep to a further study. We then used the solutions 
to explore the complex structure space of the CY manifold. Since we knew the time behavior of 
$G_{i\bar{j}} \dot z^i \dot z^{\bar{j}}$, we have solved the moduli field equation to get 
the time dependence of the complex structure moduli,
the K\"{a}hler metric, the K\"{a}hler potential and the volume of the Calabi-Yau manifold. 
That's for a Hodge number $h_{2,1}=1$, which means the dimensions of $M_C$ is unity and there is a single moduli,
since  $z^i$ are considered the coordinates that define $M_C$.  
The time dependence of The K\"{a}hler potential is negative like the K\"{a}hler metric. 
The absolute value of the potential increases with time. 
Also the volume of the CY manifold infinitely increases with time which is a deduction that 
the brane- world and the bulk are keeping expanding with time.


\end{document}